\documentclass[runningheads]{llncs}

\usepackage{graphicx}
\usepackage{url}
\usepackage{multirow}
\usepackage{tabularx}
\usepackage[english, boxruled, linesnumbered, vlined]{algorithm2e}
\usepackage{xcolor}
\usepackage{comment}
\usepackage{paralist}

\usepackage[font={small}]{caption, subfig}

\begin{document}

\title{Predicting Knowledge Gain for MOOC Video Consumption}

\author{Christian Otto \inst{1} \orcidID{0000-0003-0226-3608} \and Markos Stamatakis \inst{2}\orcidID{0000-0002-7974-308X} \and Anett Hoppe \inst{1,2}\orcidID{0000-0002-1452-9509} \and Ralph Ewerth \inst{1,2} \orcidID{0000-0003-0918-6297}}
\institute{L3S Research Center, Leibniz Universit\"at Hannover, Germany\and TIB -- Leibniz Information Centre for Science and Technology, Hannover, Germany \newline\email{first.lastname@tib.eu}}
\authorrunning{Otto et al.}
\maketitle
\begin{abstract}
Informal learning on the Web using search engines as well as more structured learning on MOOC platforms have become very popular in recent years.
As a result of the vast amount of available learning resources, intelligent retrieval and recommendation methods are indispensable -- this is true also for MOOC videos.
However, the automatic assessment of this content with regard to predicting (potential) knowledge gain has not been addressed by previous work yet.   
In this paper, we investigate whether we can predict learning success after MOOC video consumption using 1) multimodal features covering slide and speech content, and 2) a wide range of text-based features describing the content of the video. 
In a comprehensive experimental setting, we test four different classifiers and various feature subset combinations. 
We conduct a detailed feature importance analysis to gain insights in which modality benefits knowledge gain prediction the most.

\end{abstract}



\newcommand{\todo}[1]{\textcolor{red}{#1}}

%
%

\vspace{-0.3cm}
\section{Introduction}
\label{sec:introduction}
\vspace{-0.1cm}
Today's Web environment has become a valuable resource for human learning, with content available to explore an abundance of knowledge -- from sophisticated science topics like particle physics to everyday tasks such as changing a bike's tire. Especially video platforms such as YouTube gain more and more momentum in this field -- a study states that about 50\% of the daily views target some kind of learning resource \cite{smith2018}. However, with 500 hours of new content uploaded to YouTube alone every minute \cite{clement2019}, it is obvious that learners require effective search and recommendation tools to find fitting content. 

Research on automatic assessment of learning resources has targeted a number of possible dimensions, such as the prediction of user engagement towards a certain learning resource~\cite{bulathwela2020} or the correlation of knowledge gain and layout features ~\cite{shi2019investigating}. While these are interesting research directions, they do not address the question of potential usefulness of a resource for learning. This usefulness is often conceptualized as the learning success that a certain user may achieve by using the resource.

In this paper, we report on work in progress on knowledge gain prediction for lecture videos using multimodal content features.
We extend Shi et al.'s feature set~\cite{shi2019investigating} with a large number of text-based features (more than 380) and adapt them, where needed, to slide and speech content. 
In our comprehensive experiments, we investigate whether multimedia features and text-based features for slides and speech can be used or combined for knowledge gain prediction. 
We also consider that the user's capabilities might play an important role in this context.
Experiments for a set of lecture videos from the EdX platform \url{https://www.edx.org/} show that best results are obtained when the proposed text-based features are incorporated.

%
%
\vspace{-0.3cm}
\section{Related Work}
\label{sec:related_work}
\vspace{-0.1cm}
\paragraph{Analysis of (video-based) learning materials: }Navarrete et al. \cite{navarrete2021} summarize recent research in the analysis of videos as learning resources, and provide a classification of commonly analyzed features in learning videos. Those comprise features
\begin{inparaenum}
 \item extracted from the video's modalities (audio and visual features); 
 \item capturing elements of the video design (e.g., production style, usage of instructional design principles, interactive elements such as quizzes, and the instructor's behavior);
 \item related to user interaction (i.e., recorded interactions on the video such as pausing and resuming); 
 \item and, finally, textual features as they can be extracted from the visual content of the video or spoken word. 
\end{inparaenum}

\noindent As stated by the authors, the most commonly used textual resource is the video's speech transcript. 
Other textual features include the information included in the metadata (e.g., title, keywords and tags), scene text extracted from the visual channel and textual highlighting. In accordance with our own review of the literature, there is no mention of research works that perform a detailed analysis of textual features with the objective to predict a video's suitability for learning. 


\paragraph{Assessment of learning success: } Approaches assessing the success of a user's learning journey differ in their granularity (i.e., continuous monitoring of the learning process vs. testing at specific points in time) and their scalability (i.e., individual assessment vs. standardized testing). 
Research on online learning and learning during search often rely on easy-to-evaluate multiple choice tests. Most commonly, the study participants fill in a pre- and post-test; the difference in their scores is then qualified as the so-called \emph{knowledge gain}~\cite{collins2016assessing,vakkari2016,gadiraju2018,yu2018predicting,otto2021predicting}. An alternative approach was proposed by Bhattacharya \& Gwizdka~\cite{bhattacharya2019}, who collected open written statements from study participants and compare them against an expert statement. 

%
%

\vspace{-0.3cm}
\section{Dataset and Feature Extraction}
\label{sec:method}
\vspace{-0.1cm}
We use Shi et al.'s dataset~\cite{shi2019investigating} consisting of 22 lecture videos of an \textit{edX} online course called ''Globally Distributed Software Engineering`` as well as the extracted features. 
The videos, the related presentation slides (in form of PDF files) and the \textit{SubRip-Text} (SRT) speech transcript files were downloaded from EdX. 
Shi et al. employed 13 participants (10 men, 3 women, age $25.8 \pm 2.4$ years) with a computer science background and asked them to watch and assess eight or nine different videos from the course, resulting in a total sample size of $111$ learning sessions. In addition, they conducted pre- and post-knowledge tests with multiple-choice questionnaires that resulted in the knowledge gain (KG). 
Their feature set focuses on a variety of audio, visual and cross-modal features. Their \textit{linguistic} feature category differs from our linguistic features, since they investigated the spoken works only partially regarding speed and filler words, rather than examining the written and spoken text itself.
The following sections describe the feature extraction process \textit{per lecture} applied to the respective video (MP4), slides (PDF), and transcript (SRT) files yielding 387 features. 
The set of textual features comprises five subcategories: \textit{syntax}, \textit{lexical}, \textit{structure}, \textit{semantics}, and \textit{readability}. The full feature list, the full list of feature importance results and the utilized code can be found on GitHub\footnote{\scriptsize\url{https://github.com/TIBHannover/mooc_knowledge_gain}}.

\vspace{-0.3cm}
\subsection{Syntactic Features (308)}
\label{sec:syntactic}
\vspace{-0.1cm}
For the extraction of the syntactic features, we used the library \textit{Stanza}\footnote{\scriptsize\url{https://stanfordnlp.github.io/stanza/}}. It supports 66 languages, tokenization, lemmatization, and Part-of-Speech (POS) tagging. In addition, it provides an interface for the Stanford CoreNLP library~\cite{manning2014stanford} that generates syntax trees to represent the structure of a sentence.
For our experiments, we consider the full list of \textbf{word types} given by the \textit{Universal POS tags}\footnote{\scriptsize\url{https://universaldependencies.org/u/pos/})}, their average frequency per sentence in each video, once for the PDF and for the SRT files, and the ratio compared to the total number of words in the sentence per word type. Since nouns and pronouns are a majority in the given types, we record their ratio as well. Words that can not be assigned to a class are denoted as \textit{X} in the \textit{Other} class. This results in 110 calculated features.
Next, we extract $88$ \textbf{temporal features}. As reported by Kurdi et al.~\cite{Kurdi2017Lexical}, less complex texts prefer simpler tenses and thus can be used to measure text complexity. 

We utilize XPOS annotations (also generated by \textit{Stanza}) and a list of rules, that we derived from an English teaching website\footnote{\scriptsize\url{https://www.englisch-hilfen.de/grammar/englische_zeiten.htm}}. We determine and count the appearing tenses for every word of a clause. We analyze clauses individually by splitting sentences at commas or semicolons so their tenses do not interfere with each other. Lastly, we account for passive and active forms individually and also their respective frequencies compared the entirety of occurrences is considered as a feature. The next category are \textbf{phrases}, where we identify $86$ different features. 14 different phrase types were considered, extracted by \textit{Stanza's} constituency parser and defined by \textit{Penn Treebank}~\cite{Taylor2003Penn}. ''Wh-words`` are interrogatives like \textit{where, who, when}, but also \textit{how}. Similar to the previous feature types, we compute the raw count, the ratio of each individual phrase type to the entirety of occurrences, and the average frequency per sentence. Additionally, the average number of phrases per sentence is determined. Also, 24 \textbf{other syntactic features} are extracted. According to Kurdi et al.~\cite{Kurdi2017Lexical}, who found that tri-grams and tetra-grams influence the textual complexity, we count the average number of n-grams per sentence. Furthermore, the number of characters per video for slides and transcripts are computed, and in a similar fashion, the number of words as well as the minimal, average, and maximal number of words per modality. Finally, the number of expressions and questions as well as their ratio to the total number of sentences is computed as a feature.

\vspace{-0.2cm}
\subsection{Readability (12)}
\label{sec:readability}
\vspace{-0.1cm}

Stajner et al.~\cite{stajner2012what} and Brigo et al.~\cite{brigo2015clearly} suggest readability indices to measure text complexity.
Accordingly, we implement Flesch-Reading-Ease~\cite{stewart2003recalibrating}, its successor the Flesch-Kincaid and the Gunning-Fog Index~\cite{brucker2009arkansas}, the SMOG~\cite{mc1969smog}, Coleman-Liau ~\cite{coleman1975computer} as well as the Automated Readability Index (ARI)~\cite{smith1970ari}.

The input parameters for these indices involve, besides the number of letters, words, sentences, and syllables, the number of \textit{long} (4+ syllables) and \textit{difficult} words. \textit{Difficult} words have, based on Brucker's~\cite{brucker2009arkansas} definition, at least three syllables, not counting common suffixes like \textit{es, ed, ing} while not being a name or compound word. To identify these two types of words we use the \textit{CompoundWordSplitter} (\url{https://github.com/GokulVSD/FOGIndex}).

\vspace{-0.3cm}
\subsection{Lexical Features (36)}
\label{sec:lexical_features}
\vspace{-0.1cm}
The first two lexical features are related to \textbf{word frequencies}. With the intention of determining whether repetitions of important words influence the quality of a learning resource, we compute word frequencies~\cite{Kurdi2017Lexical}. This is realized by checking if the resulting word can be found in a list of $79\,672$ words given by the \textit{English Lexicon Project}~\cite{balota2007english} (\url{https://elexicon.wustl.edu/}), after filtering stopwords (via \textit{CoreNLP}) and lemmatization of plural forms. If the word exists, we gather additional metadata from this website, namely the number of syllables, the acquisition age (cf. Section~\ref{sec:lexical_features}) of the word, and the POS tags for later usage. Lastly, we divide the frequencies of each remaining word by the number of total words. 
Symbols and digits are not considered in this list and receive the frequency $0$. 
Composite words with a dash are associated with the frequencies of their respective parts.
Next, we identified six features related to the age of acquisition (AoA). This feature denotes the age the average human learns a certain word. Obviously, this can vary heavily and, thus, indicate the difficulty of a word. For reference, Kuperman et al.~\cite{kuperman2012age} created a list of $30\,000$ nouns, verbs and adjectives that we extended with a list of  $50\,000$ articles, pronouns, and inflections (\url{http://crr.ugent.be/archives/806}). Even though multiple ratings are given, we use Kuperman's AoA rating if the word is contained in the list, otherwise the average value of $10.36$ years. For our classification, we collect the earliest, average and latest reported AoA for each word. Again, we ignore digits and symbols. For the average AoA per video, we divide the sum of all average ages by the total number of words with an AoA.

Due to their important role in the context of readability, we choose to examine the \textbf{number of syllables} even further. We focus on words with one, two, polysyllabic, and the aforementioned \textit{difficult} words (cf. Section~\ref{sec:readability}). If possible, we get the number of syllables from the \textit{English Lexicon Project} (cf. Section~\ref{sec:syntactic}), otherwise, the \textit{SyllaPy} library\footnote{\scriptsize\url{https://github.com/mholtzscher/syllapy}}. It contains a (smaller) word list that is referenced if possible but is also able to compute the number of syllables for unknown ones. In theory, this method is still not a $100\%$ accurate since the pronunciation of words sometimes omits syllables, but is sufficient for our task. We compute the total number of syllables, the average number of syllables per word, the number of words containing one, two, or 3+ syllables, and the number if \textit{difficult} words. The ratio of all these measures to the total number of words is computed in order to normalize the features.
Finally, we want to investigate \textbf{word variations} in the text. \textit{Stanza's} lemmatization method allows us to analyze the variety in phrasing used by the author of the video, thus indicating a less repetitive, more vivid textual content. 
For this purpose, we create two sets, i.e., lists without duplicates. 
The first list contains all word occurrences in the text that a POS tag can be assigned to, while the second one represents the result of lemmatizing all words in the first one. 
The resulting features, for transcripts and slides, respectively, are the lengths of both lists and the ratio comparing their lengths to the total number of words in the respective file type.

\vspace{-0.3cm}
\subsection{Structural Features (24)}
\label{sec:structural_features}
\vspace{-0.1cm}

We extract structural features from the two types of input files, namely presentation slides and the SRT transcript files. 
Naturally, we have to apply different measures for the two file types since they vary strongly in their layout. For the presentation slides (PDF) we count the total number of lines in the entire presentation and the minimum, average and maximum for the (a) number of lines per slide, (b) number of words per slide, (c) number of words per line, and (d) number of letters per line. Furthermore, we count the total number of slides. The \textit{PyMuPDF} library is used to access the textual components of PDF documents. It is important to extract the content in natural reading order.

Lines in an SRT file are enumerated sections of the subtitles with individual start and end timestamps. We extract the number of subtitles, number of sentences and their collective display time. We also compare the display time with Ziefle's~\cite{ziefle1998effects} reading speed of 180 words per minute to gain insight whether it is possible to read the subtitles within the given time frame. We further record information about the subtitles by computing the minimum, average and maximum number of letters and words per sentence. We read the SRT files via Python 3's \textit{SRT} library and concatenate the individual subtitles before reassembling the underlying sentences with the \textit{Stanza} library. 

\vspace{-0.3cm}
\subsection{Semantic Features (6)}
\label{sec:semantic_features}
\vspace{-0.1cm}

Earlier approaches, as for example suggested by Stajner et al.~\cite{stajner2012what}, counted the number of possible interpretations of a word to get an idea of its complexity. Nowadays, semantic word embeddings are the state of the art to numerically represent words. We specifically aim at capturing the context of a sentence as opposed to individual words, thus, we choose \textit{sentence transformers}~\cite{reimers-2019-sentence-bert,reimers-2020-multilingual-sentence-bert} to semantically represent various parts of our textual features. From the large set of pre-trained, multilingual models we revert to the \textit{roberta-large-nli-stsb-mean-tokens} model\footnote{\scriptsize\url{https://huggingface.co/sentence-transformers/roberta-large-nli-stsb-mean-tokens}}, which achieved the highest score in the ''Semantic Textual Similarity`` benchmark. It returns embeddings with a dimension of $1024$, ignores punctuation but reacts more sensitive to changes in tenses, replacement of core nouns of a sentence or position changes of words in a sentence.
To compute the embedding representing an entire video's speech transcript as well as slide content we first encode each sentence separately and average the results afterwards, entailing three features: \textit{embed\_srt} and \textit{embed\_slide} and their distance in the embedding space. To reflect the semantic distance between the individual sentences we record the average distance between each sentence pair for both modalities. Also, we compare these two distances by capturing their difference. Due to their special role of capturing the content of textual information as opposed to the syntactic features we captured before, we will consider the sentence embedding vectors as a separate feature category \textit{EMBED = \{EMBED (slide), EMBED (srt), EMBED (both)\}} from here on.

\vspace{-0.3cm}
\subsection{User-specific Features (1)}
\label{sec:user_features}
\vspace{-0.1cm}
By design, all our extracted features are independent of the user, since they are based on the educational resource alone. However, our goal of knowledge gain prediction is also influenced by the learner's cognitive capabilities as well. For instance, some users might generally obtain better learning results after watching the video. In order to investigate the influence of user identity in our experiments, we add another feature subset, the person ID (\textit{USER} from here on). To prevent linear dependencies between these IDs, we represent them as one-hot-encoded vectors (13 dimensions).

%
%

\vspace{-0.3cm}
\section{Experimental Setup}
\label{sec:experiments}
\vspace{-0.1cm}

\begin{figure}
    \centering
    \includegraphics[width=0.85\textwidth]{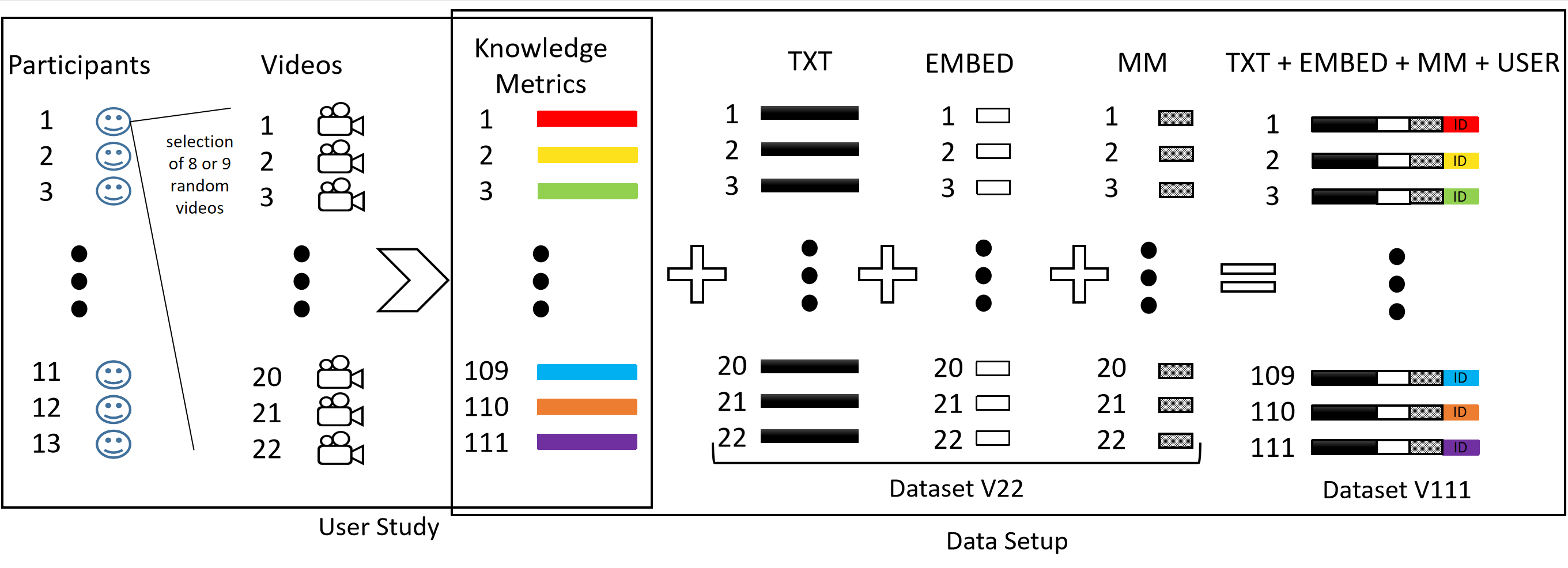}
    \vspace{-0.15cm}
    \caption{The workflow of our approach detailing the composition of our datasets for experiments \textbf{V22} and \textbf{V111} (best viewed in color).}
    \label{fig:workflow}
\end{figure}

This section describes the two knowledge gain prediction experiments that we conducted on all combinations of our feature categories. Figure~\ref{fig:workflow} gives an overview over the setup. Shi et al's user study~\cite{shi2019investigating} yielded $111$ individual learning sessions based on $13$ participants that watched eight to nine videos each. 
The extraction of features from these sessions yields, however, only $22$ unique (feature vector) samples, that is one for each video $v_i, i \in {1,..,22}$. They differ only in their target variable, the knowledge gain score $KG_i$. However, we follow Yu et al.~\cite{yu2018predicting} and do not predict knowledge gain scores directly, but rather assign one of three classes. 
We normalize the scores by transforming them into Z-scores with \textit{$\overline{X}$=$0$} and \textit{$\sigma$=$1$}. 
The knowledge gain classes are defined as follows: 1.) \textit{Low} KG, if $X < \overline{X} - \frac{\sigma}{2}$; 2.) \textit{Moderate} KG, if $\overline{X} -\frac{\sigma}{2} < X < \overline{X} + \frac{\sigma}{2}$; and 3.) \textit{High} KG, if $X > \overline{X} + \frac{\sigma}{2}$. This results in a dataset composition of 6 low, 10 moderate and 6 high for \textbf{V22} and 40 low, 40 moderate and 31 high knowledge gain samples for \textbf{V111}. For the first set of experiments (\textbf{V22}), we try to predict the average achieved knowledge gain class per participant that saw video $v_i$.
We establish a challenging \textit{knowledge gain baseline V22} by estimating the performance of participant $p_k$ on video $v_i$. Therefore, we average the knowledge gain scores of all other participants $p_l$ with $l \neq k$ who saw $v_i$ and convert it to the appropriate class afterwards, but only on videos $v \neq v_i$. 
Thus, this baseline has strong hints about the learning outcome of different participants that are not available to our classifiers. It achieves an accuracy of $45.45\%$. 
In our second set of experiments (\textbf{V111}) we add the person ID as a one-hot encoded vector to the respective video feature vectors to make them unique again, giving us the original $111$ samples. 
Target variable is the recorded knowledge gain class of the learning session. 
Again, we derive  another challenging \textit{knowledge gain baseline V111}. 
To estimate the knowledge gain class that user \textit{u} achieved on video \textit{i} we average his/her score on the $n-1$ other videos seen by him/her and, again, convert it to the appropriate class. 
This baseline is also challenging (accuracy = $43.24\%$) because the information about the user-specific learning performance is not available to our classifiers; as mentioned above, our user-specific feature is simply the encoded person-ID.

\vspace{-0.5cm}
\vspace{-1cm}

\vspace{-0.15cm}
\subsection{Data Preprocessing}
\label{sec:data_preprocessing}
\vspace{-0.1cm}

The correlation analysis of Shi et al.~\cite{shi2019investigating} (note: they did not attempt knowledge gain prediction) investigates the relationship of their multimodal features and knowledge gain. We want to examine if their features (\textit{MM} from here on) allow for knowledge gain prediction, and how our suggested features (\textit{TXT} + \textit{EMBED} + \textit{USER}) are suitable for this task, as separate feature sets and in combination. Consequently, we have seven feature combinations as inputs for experiments \textbf{V22} and \textbf{V111}: TXT, EMBED, MM, TXT+EMBED, TXT+MM, MM+EMBED, and all of them together TXT+MM+EMBED. For \textbf{V111} all of these categories also contain the one-hot-encoded person id (\textit{USER}) of the respective learner. 
We translate and scale all features 
with \textit{sklearn's} MinMaxScaler such that it is in the range between $0$ and $1$.

\subsubsection*{Dimension Reduction for Sentence Embeddings:} Since the majority of features are (single) scalars, the high-dimensional sentence embeddings ($1024$) most likely outweigh the rest. Thus, we conduct a PCA (\textit{scikit-learn}) with target dimensions of $3, 8, 16, 32$. We decide to use $16$ dimensions for the sentence embeddings since they yield $93.73\%$ explained variance for slide text, as compared to $38.26\%$ (3 dim.), $69.29\%$ (8 dim.) and $99.99\%$ (32 dim.). The results for the transcripts were similar. For the final decision, whether to use 16 or 32 dimensions, we investigated the trade-off between loss of information and accuracy in the following classification. Preliminary results showed that 16 dimensions retained better results, even though they contain around $7\%$ less information.

\subsubsection*{Data Filtering:} As a last pre-processing step we remove 47 features that are zero for every sample and thus, contain no information towards the classification. The reason is that not all occurrence-based information, e.g., tenses and word types, appear in the text.\newline

\noindent Finally, for both experiments the samples are randomly split into approximately 80\% training and 20\% test. For experiment \textbf{V111} we made sure that no video seen in training was used in test.

\vspace{-0.3cm}
\subsection{Feature Selection}
\label{sec:feature_selection}
\vspace{-0.1cm}

Breiman~\cite{breiman2001statistical} discusses the two categories of feature importance computations. The first category examines model parameters to identify what is most important towards the result, while the other one treats the model as a black box and compares simply how changing the input impacts the output. Recent work on knowledge gain prediction ~\cite{yu2018predicting,otto2021predicting} utilizes Pearson correlation to estimate the most influential features, which falls into the first category. However, Breiman describes 
the issues related to such measures as follows: 
1)~analyzing what the model does assumes that the model is the right fit for the problem, 2)~the amount of trust regarding these results is tied to the performance of the model, and 3)~this feature analysis does not tell whether the model is biased. 

Therefore, we decide to resort to a feature importance technique of the second category, namely \textit{Drop-Column Importance}, a more computational expensive type of \textit{Permutation Feature Importance}. The idea is that, for every feature \textit{f}, to train one model from scratch by dropping \textit{f}. Then, the decrease (or increase) in performance, compared to a baseline model that contains all features, shows how important \textit{f} is for the process. This technique makes the approach model-agnostic, which is useful since we are investigating multiple classifiers. The implementation we used can be found on GitHub\footnote{\scriptsize\url{https://github.com/parrt/random-forest-importances/blob/master/src/rfpimp.py}}. Negative importance values imply that the performance increases when the feature is not considered. Thus, we only keep features that have values $\geq0$ (\textbf{V22}: $40$, \textbf{V111}: $191$) and do our final run of both experiments afterwards. 
We keep the 13-dimensional person id vector for the feature selection process, since each bin represents one person and we want to investigate whether the models utilize information about the individual performances of the participants. 
We omit the PCA-transformed sentence embeddings for the importance analysis, since they are hard to interpret.
The results of the feature importance analysis are discussed in Section~\ref{sec:feature_importance}.
%
%

\vspace{-0.3cm}
\subsection{Results and Discussion}
\label{sec:results}
\vspace{-0.1cm}
We use four classifiers in our experiments, \textit{Naive Bayes (NB), Sequential Minimal Optimization (SMO), Random Forest (RF), and Multi-Layer Perceptron (MLP)} implemented by the \textit{Weka} machine learning software. For each classifier (set to default hyperparameters), each feature category, and both experiments we conduct a 5-fold cross-validation and average the results per fold in terms of precision, recall, F1-score, and accuracy. Also, for each fold, a separate feature importance analysis and feature selection is conducted to avoid bias. The following two Tables~\ref{tab:v22results} and \ref{tab:v111results} show the best performing combinations of classifier and feature category for the experiments \textbf{V22} and \textbf{V111}. 
The overall scores are macro recall, precision, and F1. 
\begin{table}[!ht]
\resizebox{\linewidth}{!}{
\begin{tabular}{|l|c|ccc|ccc|ccc|ccc|c|}
\hline
&    & \multicolumn{3}{c}{\textbf{Low}} & \multicolumn{3}{c}{\textbf{Moderate}} & \multicolumn{3}{c}{\textbf{High}} & \multicolumn{3}{c}{\textbf{Overall}} & \\ 
Feature Category & Class. & Pr & Re & F1 & Pr & Re & F1 & Pr & Re & F1 & Pr & Re & F1 & Acc. in \% \\ 
\hline \hline
Random Guess Baseline   & -     & -    & -    & -    & -    & -    & -    & -    & -    & -    & -    & -    & -    & 33.33 \\
Knowledge Gain Baseline V22 & - & 0.00 & 0.00 & 0.00 & 0.45 & 1.00 & 0.62 & 0.00 & 0.00 & 0.00 & 0.15 & 0.33 & 0.21 & 45.45 \\ 
\hline \hline
EMBED (slide)        & SMO & 0.10  & 0.20 & 0.13 & 0.45 & 0.80 & 0.57 & 0.00    & 0.00 & 0.00    & 0.18 & 0.33 & 0.23 & 42.0 \\
EMBED (srt)          & MLP & 0.10  & 0.20 & 0.13 & 0.47 & 0.50 & 0.48 & 0.30  & 0.60 & 0.40  & 0.29 & 0.43 & 0.34 & 42.0 \\
MM                   & RF  & 0.20  & 0.20 & 0.20  & 0.50  & 0.60 & 0.55 & 0.30  & 0.30 & 0.30  & 0.33 & 0.37 & 0.35 & 43.0 \\
TXT                  & NB  & 0.00    & 0.00   & 0.00    & 0.60  & 0.70 & 0.65 & 0.17 & 0.40 & 0.24 & 0.26 & 0.37 & 0.30  & 41.0 \\
MM+EMBED (slides)    & RF  & 0.20  & 0.20 & 0.20  & 0.58 & 0.70 & 0.63 & 0.07 & 0.20 & 0.10  & 0.28 & 0.37 & 0.31 & 42.0 \\
TXT+EMBED (both)     & NB  & 0.00    & 0.00   & 0.00    & 0.58 & 0.80 & 0.67 & 0.17 & 0.40 & 0.24 & 0.25 & 0.40  & 0.30 & 45.0 \\
TXT+EMBED (slide)    & NB  & 0.00    & 0.00   & 0.00    & 0.60  & 0.80 & 0.69 & 0.17 & 0.40 & 0.24 & 0.26 & 0.40  & 0.31 & 45.0 \\
MM+TXT               & RF  & 0.10  & 0.20 & 0.13 & 0.55 & 0.60 & 0.57 & 0.23 & 0.60 & 0.34 & 0.29 & 0.47 & 0.35 & \textbf{46.0} \\
MM+TXT+EMBED (both)  & NB  & 0.07 & 0.20 & 0.10  & 0.65 & 0.70 & 0.67 & 0.10  & 0.20 & 0.13 & 0.27 & 0.37 & 0.30 & 42.0 \\
MM+TXT+EMBED (slide) & NB  & 0.07 & 0.20 & 0.10  & 0.65 & 0.70 & 0.67 & 0.10  & 0.20 & 0.13 & 0.27 & 0.37 & 0.30 & 42.0 \\
\hline
\end{tabular}
}
\vspace{0.1cm}
\caption{Best results for each classifier in the \textbf{V22} experiment on the respective feature category. Knowledge gain classes Low, Moderate, and High are explained in Section~\ref{sec:data_preprocessing}.}
\label{tab:v22results}
\end{table}

In \textbf{V22}, all models clearly outperform random guessing, but only one feature combination outperforms the challenging baseline of $45.45$\% regarding overall accuracy: the combination of our textual features and Shi et al's multimedia features~\cite{shi2019investigating} with a Random Forest approach. It was able to distinguish between all three classes. In second place, our set of textual features together with the semantic sentence embeddings achieved $45\%$ accuracy with a Naive Bayes classifier. However, they failed to detect the \textit{Low} knowledge gain samples entirely. Even though the other results 
do not outperform the strong knowledge gain baseline, they are noticeably better than random guessing. In summary, \textbf{V22} indicates that a multimodal approach with a Random Forest classifier is a good approach for this problem.

\vspace{-0.5cm}
\begin{table}[!ht]
\resizebox{\linewidth}{!}{
\begin{tabular}{|l|c|ccc|ccc|ccc|ccc|c|}
\hline
&    & \multicolumn{3}{c}{\textbf{Low}} & \multicolumn{3}{c}{\textbf{Moderate}} & \multicolumn{3}{c}{\textbf{High}} & \multicolumn{3}{c}{\textbf{Overall}} & \\ 
Feature Category & Classifier & Pr & Re & F1 & Pr & Re & F1 & Pr & Re & F1 & Pr & Re & F1 & Acc. in \% \\ 
\hline \hline
Random Guess Baseline  &  - & - & -  & -  & -  & -  & -  & -  & -  & -  & -  & -  & -  & 33.33 \\
Knowledge Gain Baseline V111 & - & 0.50 & 0.18  & 0.26  & 0.39  & 0.85  & 0.54  & 0.70  & 0.23  & 0.34  & 0.53  & 0.42  & 0.38  & 43.24 \\
\hline \hline
EMBED (srt)+USER        & MLP & 0.35 & 0.33 & 0.34 & 0.48 & 0.47 & 0.47 & 0.39 & 0.56 & 0.46 & 0.41 & 0.45 & 0.42 & \textbf{44.74} \\
MM+USER                 & NB  & 0.35 & 0.39 & 0.37 & 0.44 & 0.34 & 0.39 & 0.35 & 0.50  & 0.41 & 0.38 & 0.41 & 0.39 & 39.03 \\
TXT+USER               & MLP & 0.36 & 0.29 & 0.32 & 0.31 & 0.34 & 0.32 & 0.36 & 0.35 & 0.35 & 0.34 & 0.33 & 0.33 & 34.66 \\
MM+EMBED (srt)+USER    & NB  & 0.37 & 0.33 & 0.35 & 0.45 & 0.42 & 0.44 & 0.35 & 0.47 & 0.40  & 0.39 & 0.41 & 0.40  & 39.00 \\
TXT+EMBED (slide)  + USER & MLP & 0.45 & 0.35 & 0.40  & 0.46 & 0.49 & 0.47 & 0.39 & 0.33 & 0.36 & 0.43 & 0.39 & 0.33 & 40.83 \\
MM+TXT+USER             & MLP & 0.24 & 0.26 & 0.25 & 0.48 & 0.45 & 0.47 & 0.30  & 0.39 & 0.34 & 0.34 & 0.37 & 0.35 & 36.06 \\
MM+TXT+EMBED (srt)+USER & SMO & 0.35 & 0.31 & 0.33 & 0.43 & 0.48 & 0.46 & 0.27 & 0.32 & 0.29 & 0.35 & 0.37 & 0.36 & 36.15 \\    
\hline
\end{tabular}
}
\vspace{0.1cm}
\caption{Best results for each classifier in the \textbf{V111} experiment on the respective feature category. Each experiment considered the one-hot-encoded person id \textit{USER} as an additional feature.}
\label{tab:v111results}
\vspace{-1.0cm}
\end{table}

For \textbf{V111}, the best result of $44.74\%$ has been achieved by the sentence embeddings generated from the speech transcript (EMBED (srt)) fed into an MLP. 
The model clearly outperforms random guessing, and is slightly better than the knowledge gain baseline. 
Second best performance was achieved by our textual features combined with the sentence embeddings extracted from the slide content (TXT+EMBED(slide)). These results indicate that focusing on textual features, in a syntactic as well as semantic manner, is beneficial for knowledge gain prediction when the individual is represented as a variable as well. Also, neural network based approaches scored the highest in four of the seven feature categories, perhaps due to the large number of features. The other categories fell below $40.00\%$ accuracy. However, they remained above random guessing by means of accuracy. Additionally, as the feature importance analysis will highlight, the sentence embeddings generated by the speech transcript and slides were of high importance. Therefore, neither should be neglected for this task.

In summary, experiment \textbf{V111} suggests that semantic text features that describe the content of a MOOC video, are a better choice for this task than syntactic features that objectively describe the video. In comparison with \textbf{V22} that had a slightly stronger focus on multimedia features that describe the objective quality of the video, this finding could be explained by the following: On one hand, to predict the user-independent (average) learning outcome of a MOOC video (as in \textbf{V22}) it is beneficial to consider multimodal features describing general quality aspects. On the other hand, the prediction of the individual knowledge gain (\textbf{V111}) depends on a combination of content-features and the preferences of the person itself. We tried to capture this personal influence with our one-hot-encoded person id feature. As Section~\ref{sec:feature_importance} will show, the results of the feature importance analysis underline its impact on the classification result. 

\vspace{-0.3cm}
\subsection{Feature Importance}
\label{sec:feature_importance}
\vspace{-0.1cm}

Tables~\ref{tab:feature_importancev22} and \ref{tab:feature_importancev111} show the average of the five drop-column feature importances (generated by the 5-fold cross-validation) conducted on the feature categories MM+TXT+EMBED for \textbf{V22}, and MM+TXT+EMBED+USER for \textbf{V111}.

\begin{table}[!htbp]
\begin{minipage}{0.45\linewidth}
  \centering
    \begin{tabular}{|c|l|c|}
        \hline
        Type & Feature                  & FI\\
        \hline
        MM & img\_ratio\_var                   & 0,09 \\
        MM & log\_HNR\_avg                     & 0,05 \\
        MM & average\_syllable\_duration       & 0,05 \\
        MM & coverage\_of\_slide\_content\_avg & 0,05 \\
        MM & f0\_avg                           & 0,05 \\
        MM & PVQ\_avg                          & 0,05 \\
        MM & harmonicity\_avg                  & 0,05 \\
        MM & highlight                         & 0,05 \\
        MM & jitter\_avg                       & 0,05 \\
        MM & rms\_energy\_avg                  & 0,05 \\
      \hline
    \end{tabular}
    \caption{Importance of the top-10 features of the \textbf{V22} experiment.}
    \label{tab:feature_importancev22}
\end{minipage}
\hfill
\begin{minipage}{0.45\linewidth}
  \centering
    \begin{tabular}{|c|l|c|}
        \hline
        Type & Feature                  & FI\\
        \hline
        TXT  & ratio\_VP\_sli          & 0,05 \\
        TXT  & VP\_sli                 & 0,05 \\
        USER & Person\_ID\_5           & 0,04 \\
        TXT  & PP\_sli                 & 0,04 \\
        TXT  & amount\_main\_verb\_sli & 0,04 \\
        TXT  & sum\_tok\_len\_tra      & 0,04 \\
        TXT  & avg\_adj\_sli           & 0,04 \\
        TXT  & amount\_adj\_sli        & 0,04 \\
        TXT  & ratio\_adj\_sli         & 0,04 \\
        TXT  & amount\_adpos\_sli      & 0,04 \\
      \hline
    \end{tabular}
    \caption{Importance of the top-10 features of the \textbf{V111} experiment.}
    \label{tab:feature_importancev111}
\end{minipage}
\vspace{-1cm}
\end{table}

The feature importance analyses of the two experiments show significant differences. In the experiment \textbf{V22}, the important features are dominated by the multimedia features~\cite{shi2019investigating}. From the 40 features yielding a feature importance $\geq0$ only 14 were of the textual category. This is reflected in Table~\ref{tab:v22results}, where MM+TXT achieved the best performance.

In experiment \textbf{V111} the textual features from our approach obtain the highest importance scores, with a slightly stronger focus on the slide content (''\_sli`` suffix). Out of the $191$ most important features with a value $\geq0$ the first multimedia feature has rank 50. Rank 3 is of type \textit{USER} highlighting the importance of this bin in the 13-dimensional one-hot-encoded vector. This suggests that our models identified that this learner's individual performance gave hints about the eventual learning outcome in the other videos he or she saw.

In summary, the results of the feature importance analyses do not favor a certain modality. This suggests that it is beneficial to follow a workflow of our approach, that is to initially consider a broad range of features and assess 
their importance for the classification. 
Focusing on a single modality from the start may not yield optimal results as the impact of the selected features may vary heavily depending on the target scenario. 

%
%

\vspace{-0.3cm}
\section{Conclusions}
\vspace{-0.1cm}
\label{sec:conclusion}
In this paper, we have investigated whether we can predict knowledge gain for MOOC videos based on their content. 
Therefore, we have presented an exhaustive multimodal feature analysis and analyzed the individual and combined impact of these modalities for the task of knowledge gain prediction. 
We have suggested a large set of features to represent textual content of slides and speech in lecture videos.
To the best of our knowledge, this is the first successful attempt to predict potential knowledge gain of MOOC videos based solely on their content. 
This study is clearly work in progress, but can be seen as a first step to automatically assess lecture videos solely on their content.

There are several directions for future work. First, larger studies with more users and videos have to be conducted in order to replicate the results. 
Second, the applicability to other domains has to be investigated.
Third, the incorporation of other data like user ratings and comments can be another source of information. 
Last but not least, the interplay of features from different modalities should be analyzed and modeled, e.g., based on Multimedia Learning Theory.


\subsubsection*{Acknowledgments}
This work has been partly supported by the Ministry
of Science and Education of Lower Saxony, Germany,
through the Graduate training network (Nds. Promotionsprogramm) “LernMINT (LearnSTEM):
Data-\textbf{}-assisted classroom teaching in the STEM subjects”.

\bibliographystyle{splncs04}

\bibliography{references_condensed}

\end{document}